\documentclass{aa}  

\usepackage{graphicx}
\usepackage{txfonts}
\usepackage{lineno}
\usepackage{caption}
\usepackage{mathabx}
\usepackage{color}
\usepackage[normalem]{ulem}         
\usepackage{pdflscape}
\usepackage{afterpage}

\begin{document}

\title{Thermal evolution of rocky exoplanets with a graphite outer shell}

\titlerunning{Thermal evolution of rocky exoplanets with a graphite outer shell}
\authorrunning{K. Hakim et al.}

\author{Kaustubh Hakim\inst{1,2,3,*} \and Arie van den Berg\inst{3,4} \and Allona Vazan\inst{2,5,6} \and Dennis H{\"o}ning\inst{3,7} \and Wim van Westrenen\inst{3} \and Carsten Dominik\inst{2} }

\institute{Center for Space and Habitability, University of Bern, Gesellschaftsstrasse 6, 3012 Bern, Switzerland \\ \email{ kaustubh.hakim@csh.unibe.ch}
        \and Anton Pannekoek Institute for Astronomy, University of Amsterdam, Science Park 904, 1098 XH Amsterdam, The Netherlands 
        \and Department of Earth Sciences, Vrije Universiteit, De Boelelaan 1085, 1081 HV Amsterdam, The Netherlands
        \and Department of Earth Sciences, Utrecht University, Princetonlaan 8a, 3584 CB Utrecht, The Netherlands
        \and Racah Institute of Physics, The Hebrew University, Jerusalem 91904, Israel
        \and Institute for Computational Science, Center for Theoretical Astrophysics and Cosmology, University of Z{\"u}rich, 8057 Z{\"u}rich, Switzerland
        \and Origins Center, University of Groningen, Nijenborgh 7, 9747 AG Groningen, The Netherlands}

\date{Received 17 April 2019 / Accepted 10 September 2019}

 
  \abstract
   {The presence of rocky exoplanets with a large refractory carbon inventory is predicted by chemical evolution models of protoplanetary disks of stars with photospheric C/O $>0.65$, and by models studying the radial transport of refractory carbon. High-pressure high-temperature laboratory experiments show that most of the carbon in these exoplanets differentiates into a graphite outer shell.}
   {Our aim is to evaluate the effects of a graphite outer shell on the thermal evolution of rocky exoplanets containing a metallic core and a silicate mantle.}
   {We implemented a parameterized model of mantle convection to determine the thermal evolution of rocky exoplanets with graphite layer thicknesses up to 1000~km. }
   {We find that because of the high thermal conductivity of graphite, conduction is the dominant heat transport mechanism in a graphite layer for long-term evolution ($>200$~Myr). The conductive graphite shell essentially behaves like a stagnant lid with a fixed thickness. Models of Kepler-37b (Mercury-size) and a Mars-sized exoplanet show that a planet with a graphite lid cools faster than a planet with a silicate lid, and a planet without a stagnant lid cools the fastest. A graphite lid needs to be approximately ten times thicker than a corresponding silicate lid to produce similar thermal evolution.   }
   {}


   \maketitle
%

\section{Introduction}\label{sect:introduction}

Rocky exoplanets appear to be ubiquitous around all types of planet-hosting stars in our galaxy \citep{Petigura2018}. Mass-radius relations of rocky exoplanets hint at a large variety in their composition ranging from rock-iron compositions to ice-water worlds \citep[e.g.,][]{Valencia2006,Seager2007,Wagner2011,Hakim2018a}. Other indications about their compositional diversity come from spectroscopic observations of their host stars, which show a range in photospheric elemental ratios, especially Mg/Si and C/O \citep[e.g.,][]{Bond2008,DelgadoMena2010}. Chemical evolution simulations of  refractory materials, which are the building blocks of rocky planets, in protoplanetary disks of these planet-hosting stars widen their compositional diversity even further, in particular in terms of their refractory C/O ratio \citep[e.g.,][]{Bond2010b,CarterBond2012b,Moriarty2014,Dorn2019}.

Planet-hosting stars with molar C/O {\textgreater} 0.65 (cf. C/O$_{\mathrm{Sun}}\sim0.54$ ) are capable of producing short-period rocky exoplanets abundant in carbon \citep{Moriarty2014}. Although the accuracy of photospheric C/O ratio measurements of stars in the solar neighborhood is still under debate \citep[e.g.,][]{DelgadoMena2010,Petigura2011,Nakajima2016,Brewer2016}, there is a large spread in the reported C/O ratios ranging from 0.2 to 1.6. This hints that a substantial fraction of stars still may have photospheric C/O ratios exceeding 0.65 and consequently they are likely to host carbon-enriched rocky exoplanets. Even in our solar system, refractory carbon is not rare. Graphite and diamond have been observed in ureilite parent body meteorites \citep{Nabiei2018}. Graphite is also speculated to be present on the surface of Mercury \citep{Peplowski2016}. The chemical-dynamical simulations of \citet{CarterBond2012b} accounting for giant planet migration show that rocky planets around high C/O stars can contain, in addition to iron and silicates, up to 47 wt\% carbon in weight in the form of graphite, diamond, silicon carbide, and titanium carbide. Furthermore, if radial transport of dust containing refractory carbon is efficient, carbon fractions significantly larger than observed in terrestrial planets of the solar system should be possible \citep{Klarmann2018}. 

Because pressures in planetary interiors are orders of magnitude higher than pressures in protoplanetary disks, the refractory material formed in protoplanetary disks undergoes high-pressure, high-temperature processing, thereby ensuing changes in mineralogy. Laboratory experiments show that carbon-enriched rocky exoplanets containing an iron-rich core and a silicate-rich mantle can dissolve carbon only up to an order of a percent by weight and that graphite (and diamond depending on the pressure) is the dominant carbon-bearing mineral \citep{Hakim2019}. Silicon carbide is stable only under extremely reducing conditions \citep{Hakim2018b}. Titanium carbide, even if present, is expected in small amounts because of the relatively low elemental abundance of titanium. Hence, we do not consider these carbides in the context of this study. Since graphite is 25$-$40\% lower in density than regular silicate minerals and silicate melts, graphite is expected to float on a magma ocean and consequently form an outer shell in carbon-enriched planets assuming efficient density-driven segregation \citep[e.g.,][]{Keppler2019}. 

After planet formation and differentiation, the heat locked up in rocky planetary interiors, which stems from, for example, accretion and differentiation processes, core contraction, latent heat of solidification, and radioactive decay,  is gradually released to space. Physical properties control the heat transport by convection, conduction, and radiation; these properties include the planet
radius, the interior layer thicknesses, and rock and mineral properties such
as thermal conductivity and viscosity. In rocky planets, radiation has a negligible role  to play and heat is transported mainly through conduction and convection. The contribution of convective heat transport is expressed by the Nusselt number, which increases with the vigor of thermal convection from a value of unity for purely conductive heat transport \citep{Schubert2001}.

Thermal evolution and interior dynamics in solar and extrasolar planetary bodies have been studied in detail for Earth-like silicate rock compositions \citep[e.g.,][]{Schubert1979,Spohn1991,Valencia2007c,vandenBerg2010,Hoening2016,Zhao2019}. Only a few studies have focused on the thermal evolution in planetary layers with nonsilicate mineralogies such as ice \citep[e.g.,][for icy satellites and dwarf planets]{Deschamps2001,Deschamps2014}, water and ice \citep[e.g.,][for extrasolar waterworlds]{Noack2016} and diamond \citep[e.g.,][for carbon-enriched exoplanets]{Unterborn2014}. The outermost shell determines the efficiency of heat transfer from the interior to the surface and subsequently affects the interior dynamics including the tectonic mode, volcanism, deep volatile cycles, and the presence of a magnetic field \citep[e.g.,][]{Schubert2001,Hoening2019}. Consequently, these processes have the potential to affect the habitability of the surface of a planet greatly.

The presence of graphite as an outer shell in carbon-enriched rocky exoplanets presents a unique problem and is likely to influence the planetary dynamics and habitability. In addition to its low density compared to silicate and iron-rich materials, graphite has other peculiar properties including an order of magnitude higher thermal conductivity \citep[20$-$200~W~m$^{-1}$~K$^{-1}$,][]{Tyler1953,Boylan1996,Hofmeister2014} than silicates \citep[3$-$6~W~m$^{-1}$~K$^{-1}$,][]{Kobayashi1974,Hofmeister1999}, a high melting temperature of about 4500~K at all pressures of its stability \citep{Kerley2001,Ghiringhelli2005}, and metal-like specific heat of about 700~J~kg$^{-1}$~K$^{-1}$ \citep{Boylan1996}. \citet{Unterborn2014} found that the high thermal conductivity of diamond \citep[$\sim$3000~W~m$^{-1}$~K$^{-1}$,][]{Wei1993} has a significant impact on planetary cooling; in that study they assumed diamond to be homogeneously mixed with silicates owing to their similar densities.  To our knowledge, no study has focused on the thermal evolution of low-mass planets in which carbon differentiates into a graphite shell. 

In this paper, our goal is to evaluate, to first order, the effects of a graphite outer shell on the thermal evolution of rocky exoplanets. In  Sect.~\ref{sect:method}, we describe our one-dimensional parameterized thermal evolution model applied to the main layered reservoirs in these planets. In Sect.~\ref{sect:results}, we first establish the nature of heat transport in the graphite shell. Then we quantify the effects of a conductive lid made of either graphite or silicate on top of the silicate mantle on the thermal evolution of Mars-size and Mercury-size rocky exoplanets. In Sect.~\ref{sect:conclusions}, we summarize our results and discuss the implications of our results on planets that have lids with non-graphite-like thermal conductivities and planets of different sizes. 

\section{Modeling methods}\label{sect:method}

\subsection{Interior structure}\label{sect:methodStructure}

To model the thermal evolution of a planet with multiple concentric shells, realistic values of input parameters such as the average density of each layer and surface gravity, are required (see Sect.~\ref{sect:methodEvolution}). These values are determined by computing the planetary interior structure by integrating the equation describing the hydrostatic equilibrium and Poisson's equation from the center to the surface as a function of the radial distance $r$, assuming a spherically symmetric and isotropic dependence of material properties. The equations are written as
\begin{equation}\label{eq:HEpaper4}
\frac{dP}{dr} = - \rho g, 
\end{equation} 
\begin{equation}\label{eq:PEpaper4}
\frac{dg}{dr} = 4 \pi G \rho - 2 \frac{g}{r}, 
\end{equation} where $P$ is pressure, $g$ is gravitational acceleration, $G$ is the gravitational constant, and the density $\rho(P)$ is calculated using appropriate equations of state. Since temperature has a small effect on the order of a few percent on density \citep[e.g.,][]{Hakim2018a}, we ignore the effect of temperature on material density for interior structure calculations. 

For a planet with three concentric shells and a total radius $R_{\mathrm{surf}}$ (see Fig.~\ref{fig:ThermalEvolutionSketch}), three sets of equations (\ref{eq:HEpaper4}) and (\ref{eq:PEpaper4}) need to be solved and require six boundary conditions: $P(R_{\mathrm{surf}})=0$, $g(0)=0$ and four continuity conditions for $P$ and $g$ at the two interfaces of this planet with three layers. Similarly, for a planet with two layers, two sets of equations (\ref{eq:HEpaper4}) and (\ref{eq:PEpaper4}) are solved with corresponding boundary conditions. Mass is calculated by integrating the mass-continuity equation $dm/dr = 4 \pi r^{2} \rho$.

To compute material density at a certain pressure, we implemented the equations of state of graphite \citep{Colonna2011}, MgSiO$_{3}$ \citep[enstatite for $P<25$~GPa and Mg-perovskite for $P>25$~GPa;][]{Stixrude2011}, and hcp-Fe (\citet{Fei2016} for $P<234$~GPa and \citet{Hakim2018a} for $P>234$~GPa). Comparing the equations of state of graphite, enstatite, and diamond \citep{Dewaele2008}, we verified that graphite is lower in density than enstatite and diamond by 25$-$40\% at all pressures up to the highest graphite-diamond transition pressure \citep[15~GPa;][]{Ghiringhelli2005}.

Our interior structure calculations for Mars-size and smaller exoplanets show that the material density within a particular layer varies by less than 10\%. Hence we assume constant densities for graphite, silicate, and iron layers (Table~\ref{tab:InputParameters}), which are close to our calculated volume-average densities and allow us to analyze model-independent differences in our thermal evolution calculations.

\subsection{Thermal evolution model}\label{sect:methodEvolution}

To simulate the thermal evolution of the mantle, we implemented the boundary layer theory analysis of Rayleigh-B{\'e}nard convection \citep{Turcotte1967,Stevenson1983,Schubert2001}. In this section, we first provide equations governing the boundary layer  theory and then describe the two types of model setups implemented in this paper.

\subsubsection{Boundary layer theory}\label{sect:methodBoundaryLayer}

The heat fluxes at the top and bottom of the mantle ($q_{\mathrm{man-top}}$ and $q_{\mathrm{man-bot}}$) are expressed in terms of the temperature drops across the top and bottom thermal boundary layers ($\Delta T_{\mathrm{top}}$ and $\Delta T_{\mathrm{bot}}$), and the Nusselt number Nu for the entire mantle, i.e., 

\begin{equation}\label{eq:HeatFlux}
\begin{split}
q_{\mathrm{man}-j}(t) = \mathrm{Nu}(t) \frac{k \Delta T_j(t)}{h}, \ \ \  j=\mathrm{top, bot,}
\end{split}
\end{equation} where $k$ is the thermal conductivity of the mantle (either constant or temperature-dependent; Table~\ref{tab:InputParameters}) and $h$ is the height of the mantle. The Nusselt number Nu is parameterized in terms of the Rayleigh number Ra by a power-law relation \citep{Turcotte2002}, 

\begin{equation}\label{eq:RaNu}
\begin{split}
\mathrm{Nu} = f_N \mathrm{Ra}^\beta. 
\end{split}
\end{equation} Several values between 0.19$-$0.35 have been proposed for the power-law exponent $\beta$ depending on geometry, theory, and experiments \citep[][and references therein]{Wolstencroft2009}. We assumed the classical boundary layer theory exponent $\beta=1/3$ from \citet{Turcotte1967}, which is similar to the $\beta$ for internally heated systems (0.337$\pm$0.009) from \citet{Wolstencroft2009}. We took the prefactor value $f_N = 0.164$ from \citet{Wolstencroft2009}. The Rayleigh number Ra is defined in terms of the mantle properties as 

\begin{equation}\label{eq:Ra}
\begin{split}
\mathrm{Ra}(t) = \frac{\alpha g \rho^{2} C_{P} \Delta T(t) h^{3}}{k \ \eta (T)},
\end{split}
\end{equation} where the super-adiabatic temperature difference $\Delta T$ driving the convection is the sum of the temperature drops across the top and bottom thermal boundary layers ($\Delta T = \Delta T_{\mathrm{top}} + \Delta T_{\mathrm{bot}}$), $\alpha$ is the thermal expansivity, $g$ is the gravitational acceleration, $C_{P}$ is the specific heat capacity, and $\eta(T)$ is temperature-dependent viscosity. The viscosity is given by the Arrhenius law \citep{Schubert2001},

\begin{equation}\label{eq:Visc}
\begin{split}
\eta(T) = A \exp{\left ( \frac{E}{RT} \right)},
\end{split}
\end{equation} where $A$ is the rheology prefactor, $E$ is the activation energy, and $R$ is the universal gas constant. For simplicity, we ignore the pressure-dependent $PV$ term, which is additive to the $E$ term in the Arrhenius law ($V$ is the activation volume and $P$ is the pressure). This is a reasonable approximation in view of other approximations and the limited pressure range considered. The pressure-dependent term $PV$ is small for small planets. For example, for a planet with the radius of 2500~km, $PV$ is limited to about 10\% of $E$.

\subsubsection{Mantle evolution}\label{sect:methodMantleEvolution}

To perform relevant thermal evolution calculations for carbon-enriched rocky planets, we implemented two different model setups as shown in Fig. \ref{fig:ThermalEvolutionSketch}. The temperature of the mantle ($T_{\mathrm{man}}$) assuming no heat input from the core (Fig. \ref{fig:ThermalEvolutionSketch}(a)) is given by the conservation of thermal energy \citep{Schubert2001},

\begin{equation}\label{eq:Tmantle}
\begin{split}
V_{\mathrm{man}} \rho_{\mathrm{man}} C_{P,\mathrm{man}} \frac{dT_{\mathrm{man}}}{dt} = V_{\mathrm{man}} \rho_{\mathrm{man}} H(t) - A_{\mathrm{man-top}} q_{\mathrm{man-top}}(t),
\end{split}
\end{equation} where  $H(t)=H_{0} \exp{(-t/\tau)}$  is the internal heating rate per unit mass due to the radioactive decay with a characteristic exponential decay time $\tau$ (Table \ref{tab:InputParameters}),  $q_{\mathrm{man-top}}(t)$ is the heat flux through the top of the mantle, $A_{\mathrm{man-top}}$ is the surface area of the top of the mantle, $V_{\mathrm{man}}$ is the volume of the mantle, $\rho_{\mathrm{man}}$ is the average mantle density, and $C_{P,\mathrm{man}}$ is the specific heat capacity of the mantle. The temperature contrast in Eq. (\ref{eq:HeatFlux}) for $q_{\mathrm{man-top}}(t)$ is given by $\Delta T_{\mathrm{top}} = T_{\mathrm{man}}-T_{\mathrm{surf}}$, where $T_{\mathrm{surf}}$ is the planet surface temperature.

\subsubsection{Coupled core-mantle-lid evolution}\label{sect:methodCoupledEvolution}

\begin{figure*}[!ht]
  \centering
  \medskip
  \includegraphics[width=\textwidth]{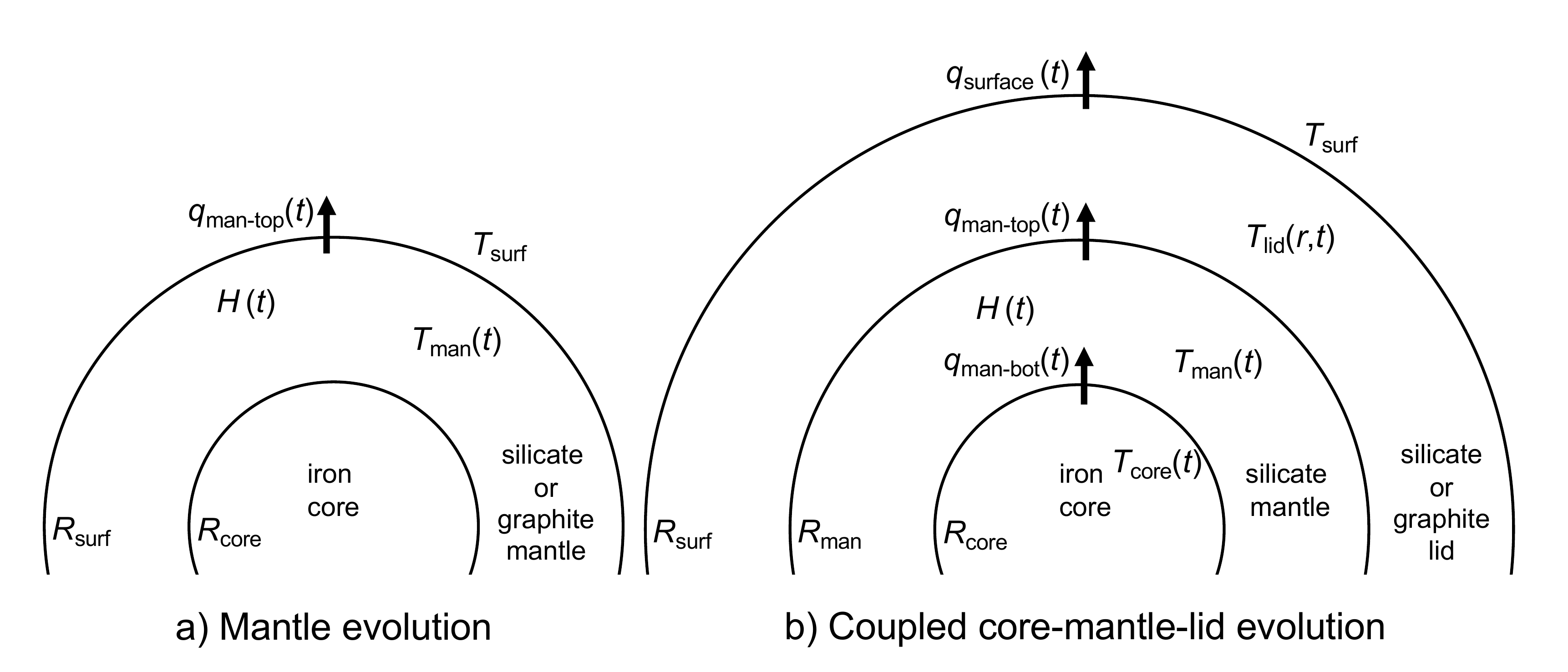}
  \caption[Thermal Evolution Sketch]{a) Mantle evolution setup (Sect.~\ref{sect:methodMantleEvolution}) for graphite and silicate mantles implemented in Sect. \ref{sect:resultsComparisonMantles}. b) Coupled core-mantle-lid evolution setup (Sect.~\ref{sect:methodCoupledEvolution}) implemented in Sects. \ref{sect:resultsComparisonLids}$-$\ref{sect:resultsKepler37b}.  }
  \label{fig:ThermalEvolutionSketch}
\end{figure*}

For models with three layers, core, mantle, and outer shell or lid, we used a coupled core-mantle-lid setup as shown in Fig. \ref{fig:ThermalEvolutionSketch}(b). The thermal evolution of the mantle coupled to that of the core is given by the conservation of thermal energy,

\begin{equation}\label{eq:CoupledTmantle}
\begin{split}
V_{\mathrm{man}} \rho_{\mathrm{man}} C_{P,\mathrm{man}} \frac{dT_{\mathrm{man}}}{dt} = V_{\mathrm{man}} \rho_{\mathrm{man}} H(t) - A_{\mathrm{man-top}} q_{\mathrm{man-top}}(t) \\ + A_{\mathrm{man-bot}} q_{\mathrm{man-bot}}(t),
\end{split}
\end{equation} where  $H(t)=H_{0} \exp{(-t/\tau)}$  is the internal heating rate per unit mass due to the radioactive decay with a characteristic exponential decay time $\tau$ (Table \ref{tab:InputParameters}),  $q_{\mathrm{man-top}}(t)$ is the heat flux through the top of the mantle, $A_{\mathrm{man-top}}$ is the area of the top of the mantle, $q_{\mathrm{man-bot}}(t)$  is the heat flux through the bottom of the mantle,  $A_{\mathrm{man-bot}}$ is the area of the bottom of the mantle, $V_{\mathrm{man}}$ is the volume of the mantle,  $\rho_{\mathrm{man}}$ is the average mantle density, and $C_{P,\mathrm{man}}$ is the specific heat capacity of the mantle. The temperature contrasts in Eq. (\ref{eq:HeatFlux}) for $q_{\mathrm{man-top}}(t)$ and $q_{\mathrm{man-bot}}(t)$ are given by $\Delta T_{\mathrm{top}} = T_{\mathrm{man}}-T_{\mathrm{lid-bot}}$ and $\Delta T_{\mathrm{bot}} = T_{\mathrm{core}}-T_{\mathrm{man}}$, respectively, where $T_{\mathrm{lid-bot}}$ is the temperature at the bottom of the lid.

The core is modeled as a heat reservoir with temperature $T_{\mathrm{core}}$ and its thermal evolution is described by another equation for the conservation of thermal energy,

\begin{equation}\label{eq:CoupledTcore}
\begin{split}
V_{\mathrm{core}} \rho_{\mathrm{core}} C_{P,\mathrm{core}} \frac{dT_{\mathrm{core}}}{dt} = - A_{\mathrm{man-bot}} q_{\mathrm{man-bot}}(t),
\end{split}
\end{equation} 
where  $V_{\mathrm{core}}$ is the volume of the core, $\rho_{\mathrm{core}}$ is the average core density, and $C_{P,\mathrm{core}}$ is the specific heat capacity of the core.

We modeled the outer shell or lid as a purely conductive static medium. Assuming a spherically symmetric temperature distribution of the lid, the partial differential equation (PDE) for time-dependent conductive heat transport \citep{Schubert2001} can be written as 

\begin{equation}
\begin{split}
  \rho_{\mathrm{lid}} C_{P,\mathrm{lid}} \frac{\partial T_{\mathrm{lid}}}{\partial t} =
  \frac{1}{r^2} \frac{\partial}{\partial r}
    \left (  r^2 k_{\mathrm{lid}} \frac{\partial T_{\mathrm{lid}}}{\partial r} \right ),
\label{eq:CoupledTlid}
\end{split}
\end{equation} where $T_{\mathrm{lid}}(r,t)$ is the lid temperature at a radial distance $r$ and time $t$, $\rho_{\mathrm{lid}}$ is the average lid density, $k_{\mathrm{lid}}$ is the thermal conductivity of the lid, and $C_{P,\mathrm{lid}}$ is the specific heat capacity of the lid. The boundary conditions applied are a prescribed fixed temperature at the outer surface of the lid ($T_{\mathrm{lid}}(R_{\mathrm{surf}},t) = T_{\mathrm{surf}}$); and a prescribed time-dependent heat flux at the interface between the lid and underlying mantle ($- k_{\mathrm{lid}} \frac{\partial T_{\mathrm{lid}}}{\partial r} = q_{\mathrm{man-top}}(t)$). The conductive lid is thermally coupled to the underlying convective mantle through the thermal boundary conditions, where the bottom heat flow is obtained from the convection model for the mantle. The time-dependent bottom temperature of the lid, on the other hand, is applied  as a boundary condition for the convecting mantle part of the domain. 

Eq.~(\ref{eq:CoupledTlid}) is solved numerically by a finite-difference discretization method using 100 grid points in the radial direction \citep{vanKan2014}. Time discretization then results in a system of algebraic equations that are solved with a time stepping algorithm that combines the solution of the conductive lid and the convecting mantle coupled through the boundary conditions.

\begin{table*}[!ht]
\caption{\label{tab:InputParameters} Input parameters for thermal evolution modeling } 
\begin{center}
\resizebox{\hsize}{!}{
\begin{tabular}{l|c|l} \hline \hline \rule[0mm]{0mm}{0mm}
Parameter & Value & Description \\[1mm]

\hline \\[-1mm]
Material properties \\
$\rho_{\mathrm{gra}}$ [kg/m$^3$]                & 2300                              & Average graphite density (Sect.~\ref{sect:methodStructure}) \\
$\rho_{\mathrm{sil}}$ [kg/m$^3$]                & 3300                              & Average enstatite density (Sect.~\ref{sect:methodStructure}) \\
$\rho_{\mathrm{iron}}$ [kg/m$^3$]               & 9000                              & Average iron density (Sect.~\ref{sect:methodStructure}) \\
$C_{P,\mathrm{gra}}$ [J K$^{-1}$ kg$^{-1}$]     &  700                              & Specific heat of graphite \citep{Boylan1996} \\
$C_{P,\mathrm{sil}}$ [J K$^{-1}$ kg$^{-1}$]     & 1250                              & Specific heat of silicate \citep{Schubert2001} \\
$C_{P,\mathrm{iron}}$ [J K$^{-1}$ kg$^{-1}$]    &  550                              & Specific heat of iron \citep{Schubert2001} \\
$\alpha_{\mathrm{gra}}$ [K$^{-1}$]              & $3\times10^{-5}$                  & Thermal expansivity of graphite \citep{Morgan1972} \\
$\alpha_{\mathrm{sil}}$ [K$^{-1}$]              & $3\times10^{-5}$                  & Thermal expansivity of silicate \citep{Schubert2001} \\
$k_{\mathrm{gra}}$($T$) [W m$^{-1}$ K$^{-1}$]   & $42327 T^{-1.035} + 0.00103 T$    & Thermal conductivity of graphite \citep[][Graphite AXM, Table 1]{Hofmeister2014} \\
$k_{\mathrm{sil}}$ [W m$^{-1}$ K$^{-1}$]        &  5                                & Thermal conductivity of silicate \citep{Schubert2001} \\
$E_{\mathrm{gra}}$ [kJ mol$^{-1}$]              & 209                               & Activation energy of graphite \citep{Wagner1959}  \\
$E_{\mathrm{sil}}$ [kJ mol$^{-1}$]              & 300                               & Activation energy of silicate \citep{Schubert2001} \\
$A_{\mathrm{gra, min.}}$ [10$^{9}$ Pa s]        & 5.3                               & Rheology prefactor for graphite \citep[min. shear modulus,][]{Cost1968} \\
$A_{\mathrm{gra, max.}}$ [10$^{9}$ Pa s]        & 185                               & Rheology prefactor for graphite \citep[max. shear modulus,][]{Min2011} \\
$A_{\mathrm{sil}}$ [10$^{9}$ Pa s]              & 160                               & Rheology prefactor for silicate (assuming $\eta$ (1600~K) = 10$^{21}$~Pa~s) \\
\\
Model properties \\
$T_{\mathrm{surf}}$ [K]                         & 700                               & Planet surface temperature \citep[Kepler-37b,][]{Barclay2013} \\
$T_{\mathrm{0,lid-bot}}$ [K]                    & 1700                              & Initial temperature at the bottom of the lid \\
$T_{\mathrm{0,man}}$ [K]                        & 2000                              & Initial mantle temperature \\
$T_{\mathrm{0,core}}$ [K]                       & 3000                              & Initial core temperature \\
$H_{0}$ [10$^{-12}$ W kg$^{-1}$]                & 34.5                              & Initial internal heating rate of the mantle \citep{Turcotte2002} \\ 
$\tau$ [Gyr]                                    & 2.95                              & Characteristic decay time of radioactive \citep{Turcotte2002} \\[1mm]
\hline
\end{tabular}
}
\end{center}\caption*{}
\end{table*}

\begin{table*}[!ht]
\caption{\label{tab:PlanetParameters} Planet parameters } 
\begin{center}
\begin{tabular}{l|ccrl} \hline \hline \rule[0mm]{0mm}{0mm}
Model & $R_{\mathrm{surf}}$ [km] & $R_{\mathrm{man}}$ [km] & $R_{\mathrm{core}}$ [km] & $g$ [m/s$^2$] \\[1mm]
\hline \\[-1mm]
Mantle evolution (100 km mantle)                    & 1600 & $-$  & 1500 & 3.7  \\
Mantle evolution (200 km mantle)                    & 1700 & $-$  & 1500 & 3.7  \\
Mantle evolution (500 km mantle)                    & 2000 & $-$  & 1500 & 3.7  \\
Mantle evolution (1000 km mantle)                   & 2500 & $-$  & 1500 & 3.7  \\
\hline
Coupled core-mantle-lid (1 km lid)                  & 3001 & 3000 & 1500 & 3.5  \\
Coupled core-mantle-lid (50 km lid)                 & 3050 & 3000 & 1500 & 3.5  \\
Coupled core-mantle-lid (500 km lid)                & 3500 & 3000 & 1500 & 3.5  \\
Coupled core-mantle-lid, Kepler-37b (1 km lid)      & 2166 & 2165 & 1083 & 2.4  \\
Coupled core-mantle-lid, Kepler-37b (100 km lid)    & 2166 & 2066 & 1083 & 2.4  \\[1mm]
\hline
\end{tabular}
\end{center}\caption*{}
\end{table*}

\subsection{Modeling assumptions}\label{sect:methodModels}

In this section, modeling assumptions for the mantle evolution and coupled core-mantle-lid setups are provided. Tables \ref{tab:InputParameters} and \ref{tab:PlanetParameters} list the material and planet properties used for modeling.

\subsubsection{Material properties}\label{sect:methodMaterial}

All relevant material properties concerning Eqs.~(\ref{eq:HeatFlux})$-$(\ref{eq:CoupledTlid}) for graphite, silicate, and iron are given in Table~\ref{tab:InputParameters}. For the viscosity of graphite, the strain rate equation from \citet{Wagner1959} is implemented, which gives the rheology prefactor $A$ in terms of the shear modulus $\mu$ and corresponding prefactor $B=1.75$ as $A = \mu /2 B$. The shear modulus of graphite has been reported to be as low as 10~GPa \citep{Cost1968} and as high as 350~GPa \citep{Min2011}. For this reason we implement two end-member rheology prefactors for graphite (Table~\ref{tab:InputParameters}).  For thermal conductivity of graphite, we used the \citet{Hofmeister2014} model with a temperature dependence given in Table~\ref{tab:InputParameters}. However, we also quantified the effect of a temperature-independent thermal conductivity of graphite in Sect.~\ref{sect:resultsComparisonMantles}.

\subsubsection{Mantle evolution setup properties}\label{sect:methodPlanetsOneLayer}

In Table~\ref{tab:PlanetParameters}, we list all models implemented in Sect.~\ref{sect:results}. The mantle evolution setup (Fig. \ref{fig:ThermalEvolutionSketch}(a)) is used in Sect.~\ref{sect:resultsComparisonMantles} to illustrate that a graphite mantle exits the convection regime of heat transport and enters the conductive regime much earlier than a silicate mantle. For this purpose, we defined the duration of convective cooling as the time required for the Nusselt number to reach unity. We integrated Eq.~(\ref{eq:Tmantle}) to compute the mantle temperature evolution supplemented by Eqs.~(\ref{eq:HeatFlux})$-$(\ref{eq:Visc}) and parameter values for graphite or silicate from Table~\ref{tab:InputParameters}. We kept the core size of our model fixed at 1500~km and the mantle thickness the same for the graphite and silicate cases (see Table~\ref{tab:PlanetParameters}). This setup allowed us to isolate the effects of planet properties such as the planet size, surface area, and gravity or internal heating and initial temperature on our model outcomes. For the base case, we assumed a surface temperature of 700~K (see Table~\ref{tab:InputParameters}), identical initial mantle temperatures of 2000~K, and no radiogenic heating. To quantify the effects of initial mantle temperature, radiogenic heating, and thermal conductivity model of graphite, we varied these parameters one by one (see Sect.~\ref{sect:resultsComparisonMantles}).

\subsubsection{Coupled core-mantle-lid evolution setup properties}\label{sect:methodPlanetsThreeLayer}

In Sects. \ref{sect:resultsComparisonLids}, \ref{sect:resultsGraphiteLids}, and \ref{sect:resultsKepler37b}, we implemented the coupled core-mantle-lid evolution setup (Fig.~\ref{fig:ThermalEvolutionSketch}(b)). We assumed the core to be made of iron, the mantle to contain silicates, and the lid (if present) to be either silicate or graphite. We also implemented reference cases with extremely thin lid to simulate lidless planets in Sects. \ref{sect:resultsGraphiteLids} and \ref{sect:resultsKepler37b}. We integrated Eqs.~(\ref{eq:CoupledTmantle}), (\ref{eq:CoupledTcore}), and (\ref{eq:CoupledTlid}) for the three layers (Fig. \ref{fig:ThermalEvolutionSketch}(b)) to calculate the thermal evolution. To isolate model-dependent effects, we fixed the core and mantle radii at 1500~km and 3000~km, respectively, and only varied the lid thickness (Table~\ref{tab:PlanetParameters}). Across different models, we also assumed the same surface temperature, the same initial temperatures for the core, mantle, and lid, and the same heating rate for the mantle (see Model properties in Table~\ref{tab:InputParameters}). We assumed the internal heating in the core and lid to be zero. In Sect. \ref{sect:resultsKepler37b}, we implemented this setup to Kepler-37b with a known radius of 2166~km \citep{Stassun2017}. We fixed the core radius of our Kepler-37b models to half of the total radius and mantle and the lid thicknesses varied depending on the model (Table~\ref{tab:PlanetParameters}). 

\section{Results}\label{sect:results}

\subsection{Duration of convective cooling in graphite and silicate mantles}\label{sect:resultsComparisonMantles}

\begin{figure*}[!ht]
  \centering
  \medskip
  \includegraphics[width=0.6\linewidth]{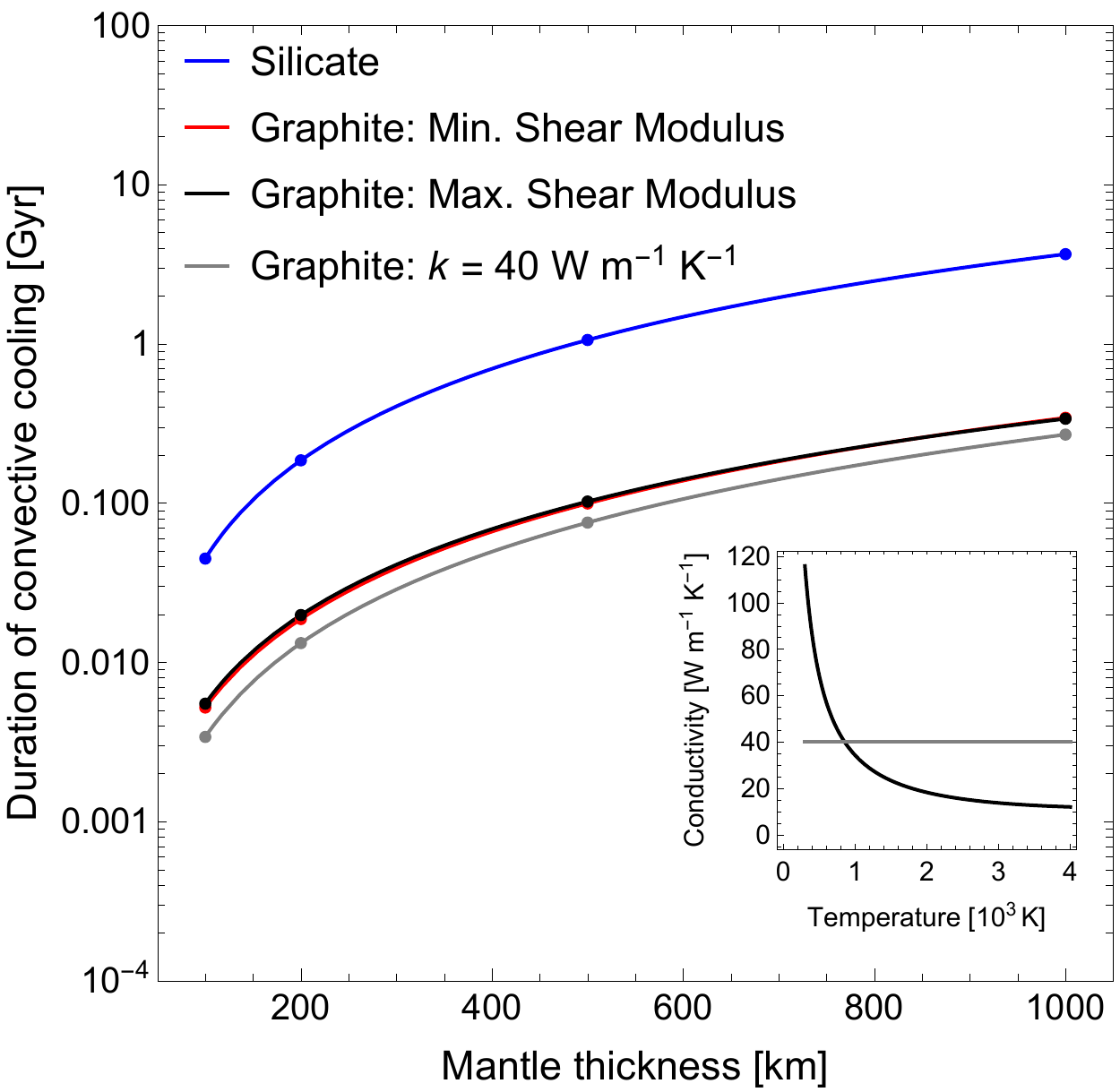}
  \caption[Duration of convection cooling]{Comparison of the convective cooling duration  of graphite and silicate mantles with a core radius of 1500~km and a mantle thickness of 100$-$1000~km. The inset panel shows the two models \citep[][and constant thermal conductivity]{Hofmeister2014} for thermal conductivity of graphite used to calculate the duration of convective cooling. For models with a graphite mantle, in addition to constant thermal conductivity, two cases of viscosity based on the minimum and maximum values of shear modulus are also compared.}
  \label{fig:DurationConvectiveCooling}
\end{figure*}

Our calculations implementing the mantle evolution setup (no lid) in Fig.~\ref{fig:ThermalEvolutionSketch}(a) show that the duration of convective cooling (see Sect. \ref{sect:methodPlanetsOneLayer}) for silicate-mantle planets with mantle thicknesses between 100$-$1000~km is between 0.05$-$3.7~Gyr (Fig.~\ref{fig:DurationConvectiveCooling}). In contrast, the convective cooling duration for graphite-mantle planets is an order of magnitude lower (0.006$-$0.34~Gyr). For graphite-mantle planets, the cases of minimum and maximum shear modulus (see Sect.~\ref{sect:methodPlanetsOneLayer}) differ by 0.3~Myr (for the 100~km case) and 4.6~Myr (for the 1000~km case), implying a negligible effect of shear modulus on the cooling of the planet. If we adopt a constant thermal conductivity of graphite (40 W m$^{-1}$ K$^{-1}$) instead of the \citet{Hofmeister2014} model, the duration of convective cooling decreases by 20$-$35\%. This is because the Hofmeister thermal conductivity is lower than 40 W m$^{-1}$ K$^{-1}$ at initial graphite mantle temperature considered in this work (see inset Fig.~\ref{fig:DurationConvectiveCooling}).  

Assuming an initial mantle temperature of 4000~K instead of 2000~K increases the lifetime of convection for the silicate and graphite cases by only 10$-$80~Myr and 0.7$-$5~Myr, respectively. Incorporating internal heating (see Table \ref{tab:InputParameters}), we find that silicate-mantle models need only up to 0.2\% more time to reach Nu$=$1 compared to the models without internal heating. As the main radiogenic heat producing elements in rocky planets (U, Th, and K) are highly incompatible in graphite, it is unlikely that significant internal heating in a graphite layer would occur under any circumstances. Even if radiogenic heating in the graphite mantle is made equal to that in silicates, the duration of convective cooling changes by less than 0.1\%. 

Although we ignore the pressure-dependent term in viscosity in our modeling, we extend our calculations to larger planets up to the size of Earth. Our calculations show that for large planets the duration of convective cooling increases by less than 20\% compared to the planets shown in Fig.~\ref{fig:DurationConvectiveCooling} for layer thicknesses up to a few hundred kilometers. Fast cooling of a graphite layer is attributed to the high thermal conductivity of graphite. 

\subsection{Thermal evolution of planets with graphite and silicate lids}\label{sect:resultsComparisonLids}

Because of its high thermal conductivity and efficient cooling, a physically separate graphite outer shell inevitably acts as an insulating stagnant lid on top of a silicate mantle. In contrast, a silicate lid may form on top of the convecting mantle as a consequence of the temperature-dependence of viscosity. The thickness of the silicate lid depends on the thermal state of the planet and increases as the planet cools. For a hot mantle and/or a large carbon inventory, the graphite outer shell could be much thicker than what the silicate lid would be. In particular planets with plate tectonics, such as Earth, do not exhibit a stagnant lid. To evaluate the effect of an outer graphite shell on the cooling rate, in this section we first compare planets with a fixed lid thickness made of either graphite or silicate. In a second step, we compare planets with different graphite lid thicknesses with each other (Sect.~\ref{sect:resultsGraphiteLids}). Finally, we compare the thermal evolution of Kepler-37b assuming a graphite lid or a silicate lid or no lid (Sect.~\ref{sect:resultsKepler37b}). 

\begin{figure*}[!htp]
  \centering
  \medskip
  \includegraphics[height=.9\textheight]{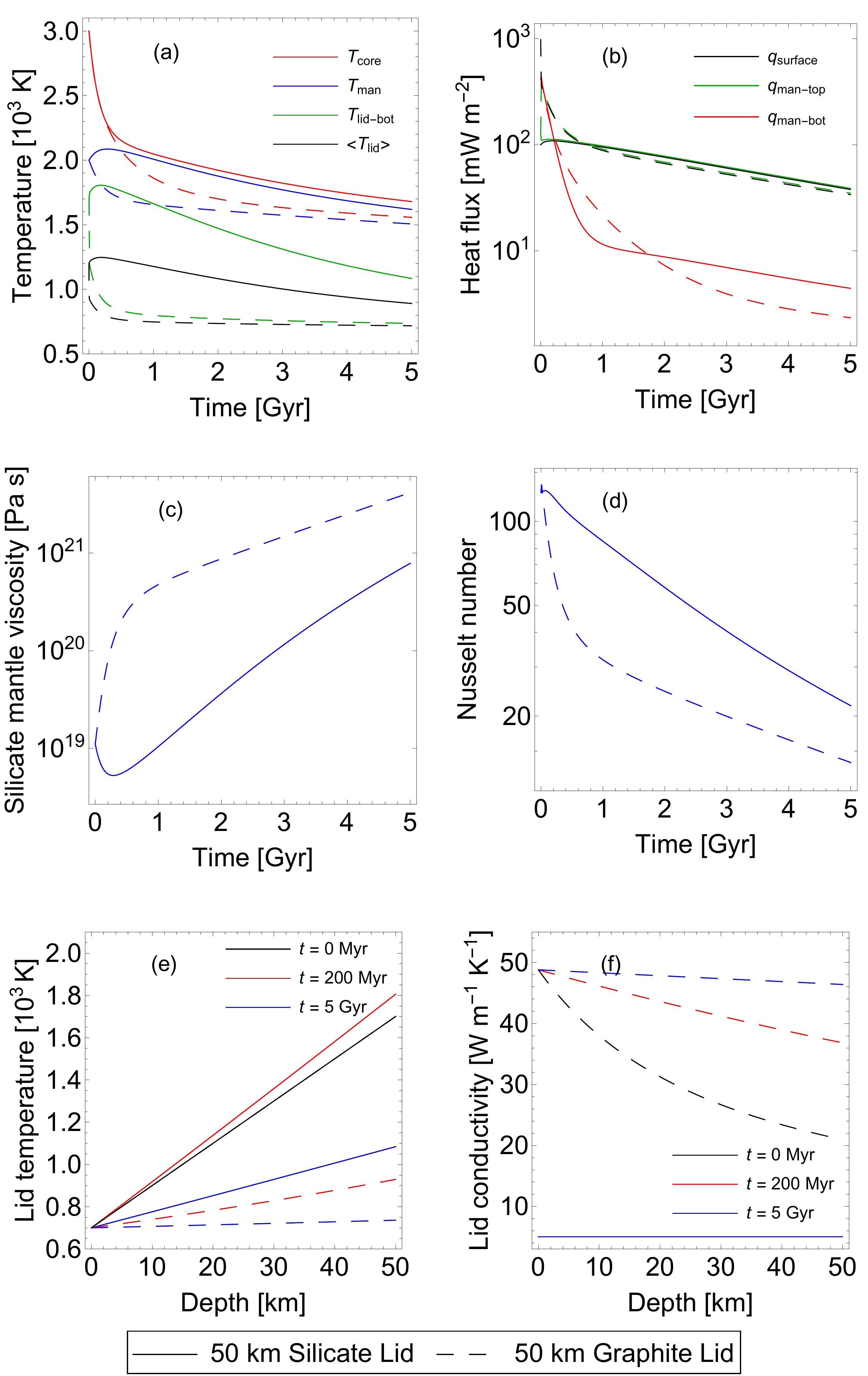}
  \caption[Thermal evolution of Kepler-37b]{Plots (a$-$f) illustrating the coupled core-mantle-lid thermal evolution for models with either a graphite or silicate lid. The core, mantle, and planetary radii are identical for the two models. The 0~Myr lines overlap in plot (e).}
  \label{fig:EvolutionSilicateLidvsGraphiteLid}
\end{figure*}

\begin{figure*}[!htp]
  \centering
  \medskip
  \includegraphics[width=0.95\linewidth]{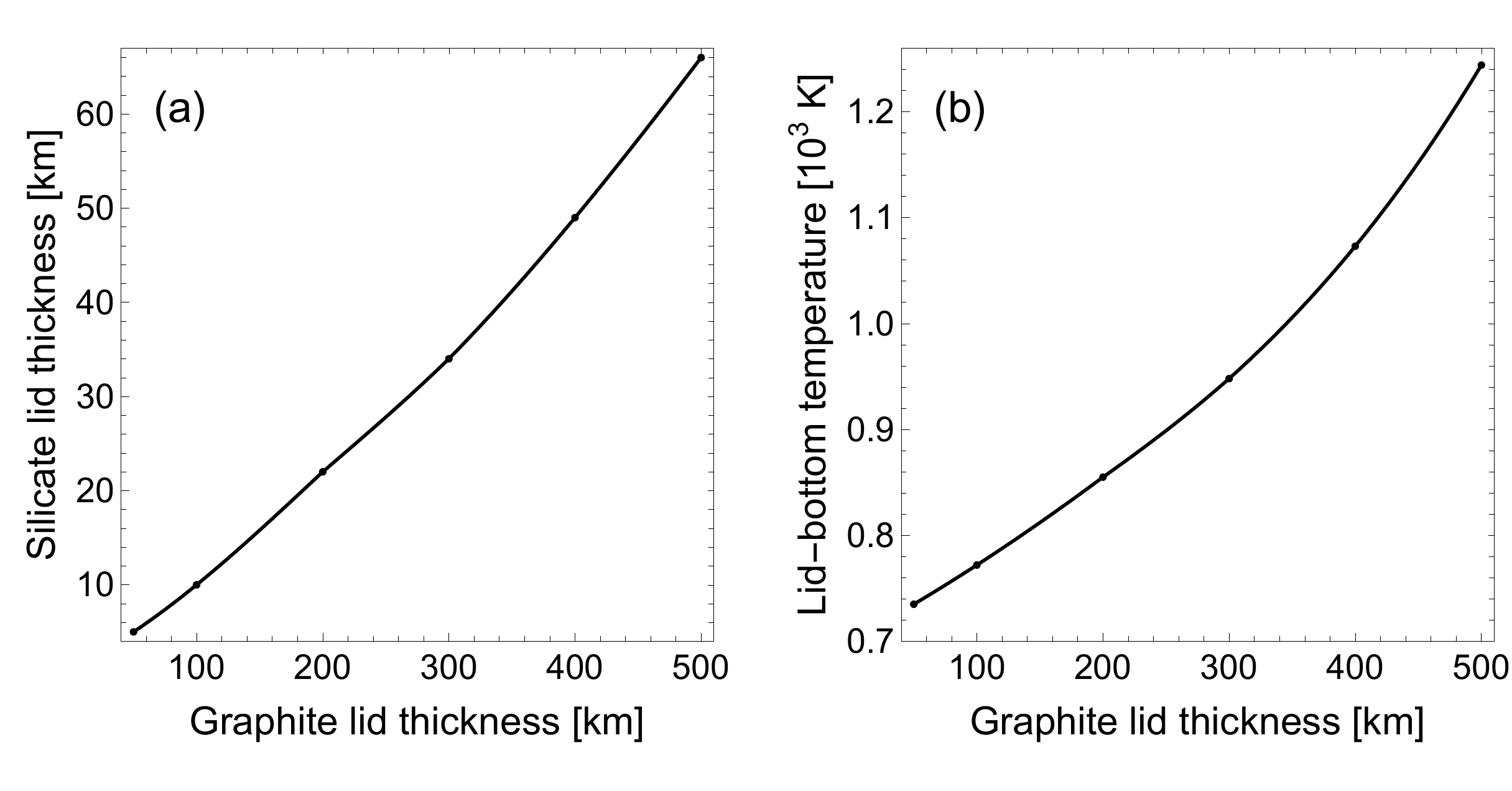}
  \caption[Silicate Lid vs Graphite Lid]{(a) The thicknesses of silicate and graphite lids required to reach the same temperature at the bottom of the lid after 5~Gyr of evolution. (b) The corresponding temperature at the bottom of the lid after 5~Gyr of evolution. The initial lid-bottom temperature is 1700~K in all cases. }
  \label{fig:SilicateLidvsGraphiteLid}
\end{figure*}


Implementing the coupled core-mantle-lid setup (Fig.~\ref{fig:ThermalEvolutionSketch}b), we compare the thermal evolution of planets with either a graphite lid or a silicate lid and a lid thickness of 50~km. See Tables~\ref{tab:InputParameters} and \ref{tab:PlanetParameters} and Sect.~\ref{sect:methodPlanetsOneLayer} for material properties and modeling assumptions. The iron core and silicate mantle radii are fixed at 1500~km and 3000~km. The internal heating rate is the same for both models. It is well known that the presence of a stagnant lid on top of a convective mantle delays the cooling of the mantle. We are interested in the differences in planetary cooling due to different lid compositions. 

Fig.~\ref{fig:EvolutionSilicateLidvsGraphiteLid} compares several properties related to planetary thermal evolution spanning 5~Gyr. Despite the same initial temperatures for both models, there is a significant difference in the evolution of temperature (Fig.~\ref{fig:EvolutionSilicateLidvsGraphiteLid}(a)). For the two cases the temperature at the bottom of the lid differs by almost 400~K and the core and mantle temperatures differ by more than 100~K. These differences between the two models are attributed to up to an order of magnitude difference in the thermal conductivity of graphite and silicate lids (Fig.~\ref{fig:EvolutionSilicateLidvsGraphiteLid}(f)). The initial thermal conductivity distribution within the graphite lid varies between 20$-$50~W~m$^{-1}$~K$^{-1}$ because of the large distribution (a difference of 1000~K between the top and bottom of the lid) in the initially assumed lid temperature profile (Fig.~\ref{fig:EvolutionSilicateLidvsGraphiteLid}(e)). Although the initial temperature distribution within the lid is the same for silicate and graphite lids, the higher thermal conductivity of graphite cools the graphite lid faster than the silicate lid. A drop in the temperature of graphite increases its thermal conductivity and makes its thermal conductivity distribution in the lid flatter (see 200~Myr and 5~Gyr profiles in Fig.~\ref{fig:EvolutionSilicateLidvsGraphiteLid}(f)). This is a direct consequence of the inverse temperature proportionality of thermal conductivity of graphite in the Hofmeister model (inset Fig.~\ref{fig:DurationConvectiveCooling}). This increased thermal conductivity of graphite lid further accelerates cooling of the lid.

The lower the temperature at the bottom of the lid, the higher is the temperature contrast across the thermal boundary layer at the top of the mantle. In the graphite lid case, this higher temperature contrast allows for a higher heat flux through the top of the mantle especially in the first 600~Myr (see Fig.~\ref{fig:EvolutionSilicateLidvsGraphiteLid}(b)). Consequently, higher heat flux allows the mantle and the core to cool faster. We note that between 0.2$-$1.8~Gyr the heat flux in the silicate lid case at the bottom of the mantle is lower than that in the graphite lid case (Fig.~\ref{fig:EvolutionSilicateLidvsGraphiteLid}(b)) because of a lower temperature contrast between the core and mantle temperature at the bottom thermal boundary layer (Fig.~\ref{fig:EvolutionSilicateLidvsGraphiteLid}(a)). As mantle viscosity is a function of the mantle temperature, it increases rapidly with time for the graphite lid case compared to the silicate lid case as seen in Fig.~\ref{fig:EvolutionSilicateLidvsGraphiteLid}(c). The minimum in the mantle viscosity in the silicate lid case at about 0.3~Gyr arises from the corresponding maximum in the mantle temperature in Fig.~\ref{fig:EvolutionSilicateLidvsGraphiteLid}(a). The Nusselt number also decreases faster for the graphite lid case than for the silicate lid case (Fig.~\ref{fig:EvolutionSilicateLidvsGraphiteLid}(d)). 

Clearly, a 50~km silicate lid significantly delays the cooling of a planet compared to a 50~km graphite lid. Another relevant comparison between silicate and graphite lids is to quantify the equivalent thickness of a silicate lid to achieve the same cooling as a graphite lid. In Fig.~\ref{fig:SilicateLidvsGraphiteLid}(a), each data point represents two models: one with a graphite lid and another with a silicate lid, which have the same temperature at the bottom of the lid after 5~Gyr of evolution. We plot this lid-bottom temperature in  Fig.~\ref{fig:SilicateLidvsGraphiteLid}(b). A silicate lid with approximately an order of magnitude lower thickness than the graphite lid is sufficient to reproduce the same temperature at the bottom of the lid after 5~Gyr. This is a significant result because it implies that a planet with a graphite lid cools similar to a planet with a silicate lid that is approximately ten times thinner. 

\subsection{Effects of graphite lid thickness on thermal evolution}\label{sect:resultsGraphiteLids}

\begin{figure*}[!htp]
  \centering
  \medskip
  \includegraphics[height=.9\textheight]{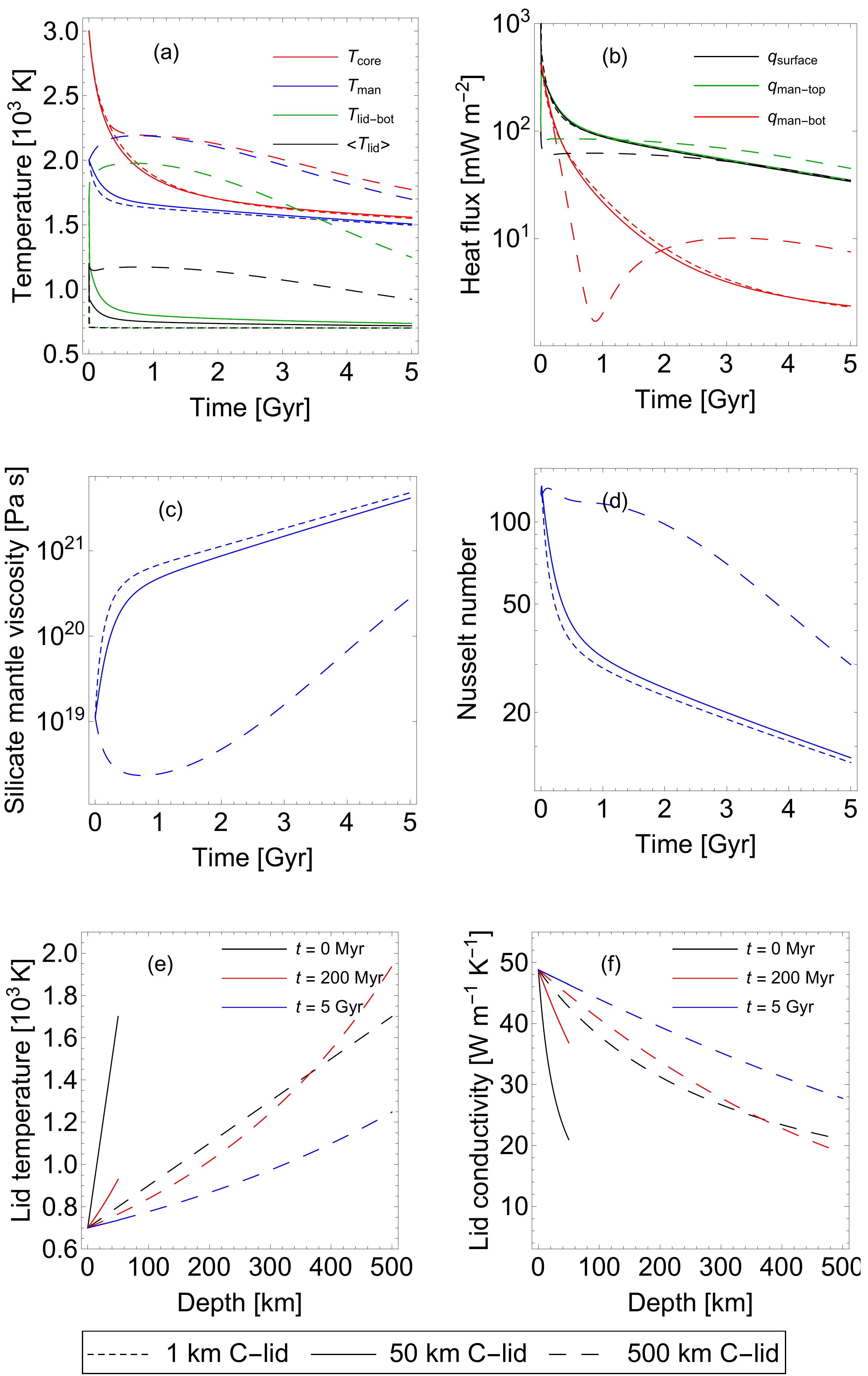}
  \caption[Thermal evolution of Kepler-37b]{Parameters related to the thermal evolution of three-layer planets (a$-$f) with core and mantle radii of 1500~km and 3000~km, respectively. All three models have different graphite lid thicknesses (1~km, 50~km, and 500~km) and consequently different planetary radii. In plots (e) and (f), the 5 Gyr lines for the 50~km and 500~km cases overlap; the 1~km case is not shown.}
  \label{fig:GenericEvolution}
\end{figure*}

In Sect.~\ref{sect:resultsComparisonLids}, we show that silicate lids are significantly more inefficient at planetary cooling than graphite lids. In this section, we model the thermal evolution of planets that do not form a silicate lid. We quantify the effect of graphite lid thickness on thermal evolution by implementing the coupled core-mantle-lid setup. We fix the core and mantle radii at 1500~km and 3000~km, and add graphite lids with thicknesses of 1~km, 50~km, and 500~km on top of the mantle. The 1~km case is introduced to emulate a planet without a conductive lid and to remove any model dependences such as the temperature contrast at the top of the mantle. Again the total internal heating is the same as it depends on the volume of the mantle, which is the same for all models. 

Fig.~\ref{fig:GenericEvolution}(a) shows that the silicate mantle and graphite lid of the 50~km model cool slower than the 1~km model because of several effects. First, the thermal inertia of the graphite lid, which is related to its heat capacity and its thermal conductivity, smoothens a rapid temperature drop in the early stages. Second, the graphite lid presents a thermal resistance that reduces the surface heat flux for a given temperature contrast \citep[see][]{vandenBerg2005}. Third, the presence of an outer shell reduces the temperature contrast at the top of the mantle, which drives thermal convection. Compared to the 50~km C-lid, the 500~km C-lid provides both a much larger thermal inertia and thermal resistance (hereafter, collectively termed as thermal shielding) resulting in much longer cooling times for its silicate mantle. Although the lid-bottom temperature and the silicate mantle temperature of the 500~km model also tend to approach the respective temperatures of the reference model after 5~Gyr, they are still higher than the other two models by about 500~K and 200~K at 5~Gyr, respectively (Fig.~\ref{fig:GenericEvolution}(a)). Our calculations for 100$-$500~km C-lid cases indicate that a thin graphite shell (\textless 200~km) exhibits small thermal shielding and does not significantly affect the long-term thermal evolution.

In Fig.~\ref{fig:GenericEvolution}(b), the heat flux at the bottom of the mantle for the 50~km case is smaller than that of the 1~km case between 0.5$-$4~Gyr. It is consistent with the smaller core-mantle temperature contrast for the 50~km case shown in Fig.~\ref{fig:GenericEvolution}(a). The smaller drop of the core-mantle temperature contrast corresponds to a steeper drop in the core temperature in the first 1~Gyr combined with a smoother drop in the temperature of the lid, which is related to the latter. In contrast, the 1~km case shows a higher core-mantle temperature contrast in line with the absence of the thermal inertia of the C-lid. For the 500~km lid model, during the first 0.4~Gyr the drop in the core temperature is similar to that of the 50~km case. However, after 0.4~Gyr, the core temperature decreases slowly as a consequence of the thermal shielding effect of the thick graphite lid, not allowing the heat to escape from the core efficiently. This results in a higher core temperature (by 200~K) for the 500~km lid model at 5~Gyr than in the other two models. The minimum in the core heat flux for the 500~km case at 0.9~Gyr corresponds to a maximum of the mantle temperature and almost disappearing temperature contrast at the core-mantle boundary. We could speculate that such an event might lead to the demise of an early planetary magnetic field through a shutdown of a core-dynamo process \citep[e.g.,][]{Stevenson2001}. 

The trends in the silicate mantle temperature corroborate the trends in the silicate mantle viscosity and the Nusselt number as seen in Fig.~\ref{fig:GenericEvolution}(c,d). Fig.~\ref{fig:GenericEvolution}(e) shows three snapshots of the radial temperature profile in the graphite lid. For the 50~km C-lid model, the temperature profile at 5~Gyr is not as steep as at 200~Myr as expected from the evolution of temperature at the bottom of the lid. For the 500~km model, since the lid temperature increases first and then decreases, its 200~Myr temperature profile crosses the initial profile. This effect is again a result of the large thermal shielding provided by the 500~km C-lid. Similar trends are observed for the radial distribution of thermal conductivity in the graphite shell (Fig.~\ref{fig:GenericEvolution}(f)).

\subsection{Application to Kepler-37b}\label{sect:resultsKepler37b}

We now apply the coupled core-mantle-lid setup (Fig.~\ref{fig:ThermalEvolutionSketch}(b)) to Kepler-37b to demonstrate differences in the thermal evolution of lidless, graphite-lid, and silicate-lid cases. For this purpose, we use three models, 1~km graphite-lid (emulating a planet without any lid), 100~km graphite-lid, and 100~km silicate-lid, respectively (Table~\ref{tab:PlanetParameters}). The core radius is fixed to half of the total radius. Although the thickness of a stagnant lid evolves with time, in this case we use models with a fixed stagnant lid thickness. This is required for a model-independent comparison between graphite and silicate lids.

\begin{figure*}[!htp]
  \centering
  \medskip
  \includegraphics[height=.95\textheight]{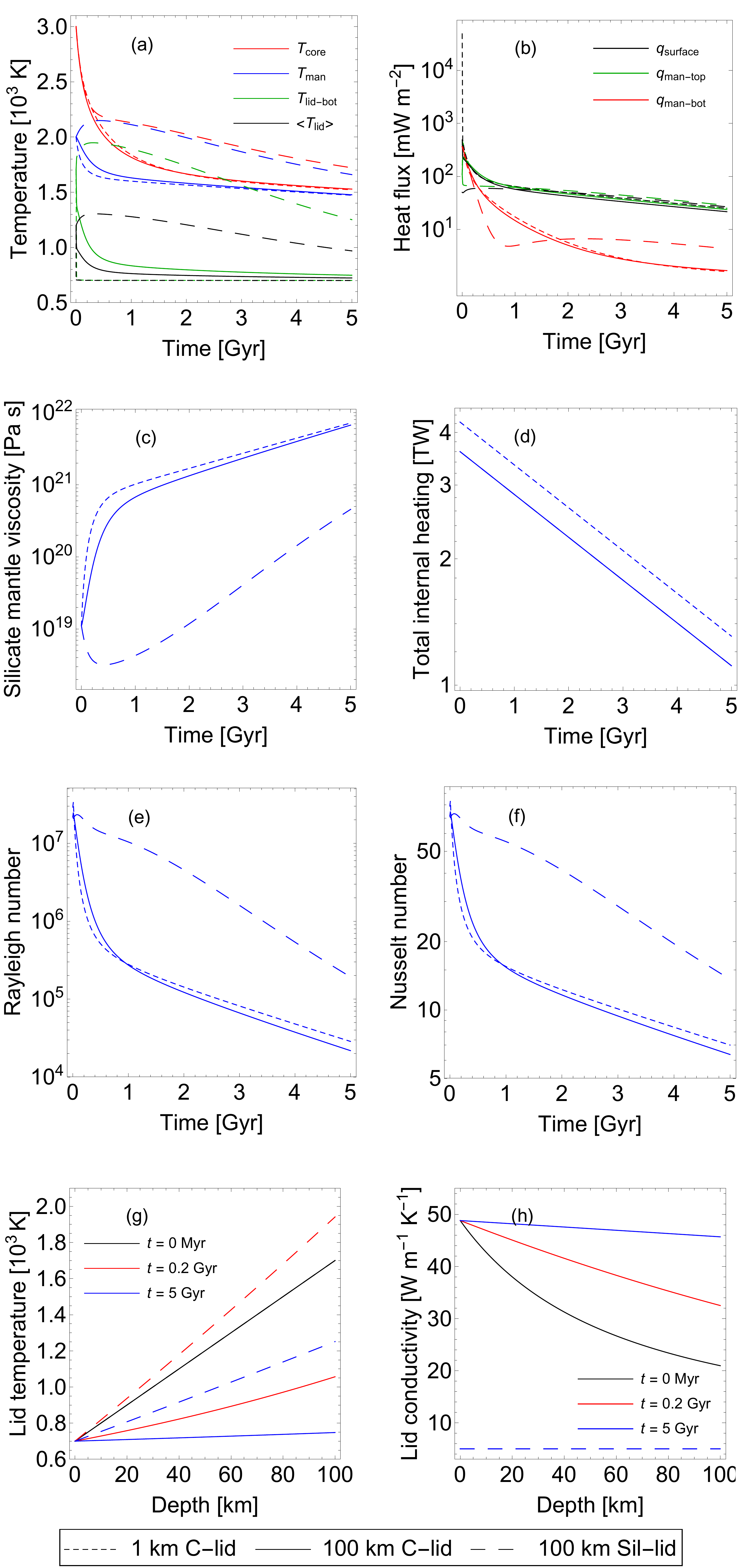}
  \caption[Thermal evolution of Kepler-37b]{Thermal evolution parameters of Kepler-37b (a$-$h) for models with 100~km thick silicate and graphite lids compared to a 1~km graphite-lid model. In plot (g) the 0 Myr lines overlap; the 1~km case is not shown in plots (g) and (h).  }
  \label{fig:Kepler37bEvolution}
\end{figure*}

\afterpage{
}

Qualitatively, the three cases shown in Fig.~\ref{fig:Kepler37bEvolution} are similar to the three cases described in Sect.~\ref{sect:resultsGraphiteLids}. As expected, a lidless planet cools the fastest, followed by the 100~km graphite-lid model, whereas the silicate-lid Kepler-37b model significantly slows down cooling, essentially behaving like a much thicker graphite-lid model as demonstrated in Fig.~\ref{fig:GenericEvolution}. The core temperature of the 100~km C-lid case in Fig.~\ref{fig:Kepler37bEvolution}(a) is close to that of the 1~km C-lid case. For the 100~km silicate lid case, the core temperature stays above the temperatures of the other two cases because of intense thermal shielding. The core heat flux in Fig.~\ref{fig:Kepler37bEvolution}(b) is smallest for the 100~km silicate lid case in the first 1.8~Gyr in line with the small core-mantle temperature contrast and an initially increasing mantle temperature. Since all radiogenic heating in the model is assumed to be concentrated in the mantle silicates and specified at the same initial value per unit mass (Table~\ref{tab:InputParameters}), the amount of internal heating decreases with the increasing thickness of the lid. That is why the total internal heating in the 100~km silicate and graphite lid models is the same but is about 14\% lower initially than the 1~km C-lid model Fig.~\ref{fig:Kepler37bEvolution}(d). However, this difference in internal heating has a small effect that is not discernible between the 1~km and 100~km C-lid models.

Silicate mantle viscosities are lower for models with higher thermal shielding (Fig.~\ref{fig:Kepler37bEvolution}(c)). Because of the direct dependence of viscosity on the mantle temperature, a local minimum is observed in the 100~km silicate-lid case owing to a corresponding temperature maximum. The Rayleigh and Nusselt numbers of the 100~km C-lid model end up higher than those of the 1~km C-lid model because of their strong dependence on the mantle thickness, which is smaller for the 100~km C-lid model (Fig.~\ref{fig:Kepler37bEvolution}(e,f)). On the other hand, the Rayleigh and Nusselt numbers of the 100~km silicate-lid models are higher than those of the other two models because of the difference in their viscosity. The trends in the radial distribution of temperature and thermal conductivity (Fig.~\ref{fig:Kepler37bEvolution}(g,h)) in the lid are similar to those in Sect.~\ref{sect:resultsGraphiteLids}.

\section{Discussion and conclusions}\label{sect:conclusions}

In this paper, we model the thermal evolution of rocky exoplanets whose chemical composition and physical structure are different from those of the terrestrial planets we know of. Not surprisingly, the thermal structure depends on the mineralogy of different layers in the planet. Carbon-enriched rocky exoplanets are expected to contain an iron core, a silicate mantle, and a graphite outer shell \citep{Hakim2019}. Our calculations show that a graphite layer is largely conductive in nature during all but the earliest stages of planetary evolution, essentially behaving like a stagnant lid with a fixed thickness. This is mainly a result of thermal conductivity of graphite being approximately one order of magnitude higher than that of common mantle silicate minerals. For the same reason, a conductive silicate lid would slow down cooling by as much as an order of magnitude thicker conductive graphite lid would do. As such our models are applicable to stagnant lid planets with different lid thermal conductivities. For example, if Mercury has a stagnant lid partially consisting of graphite in addition to silicates, its cooling might have been accelerated compared to the assumption of fully silicate lid. On the other hand, for a Mars-size planet, a 100~km lid model with half the thermal conductivity of graphite would end up with a 100~K higher temperature at the bottom of the lid at 5~Gyr than the 100~km graphite-lid model. Whereas, a 100~km model with a diamond-like thermal conductivity would cool as fast as the 1~km graphite-lid model.

As opposed to a planet without any stagnant lid, a graphite lid slows down the cooling of the planet by thermally shielding the interior due to the thermal inertia and thermal resistance of the graphite lid. The thermal inertia is mostly important during the first $\sim$100~Myr of planetary evolution when the thermal profile of the lid changes fast. The thermal resistance of the graphite lid \citep[e.g.,][]{vandenBerg2005} controls the long-term thermal evolution. We find that a thin outer graphite shell (\textless 200~km) has a small effect on the heat release from the deep interior of the planet. This implies that a thin graphite lid on top of the silicate mantle does not significantly impact the long-term evolution of the interior. However, for planets with higher graphite layer thicknesses, the thermal shielding effect of the lid becomes significant enough to slow down the cooling of the planet by several billion years. With the application to Kepler-37b, we show that a lidless model cools faster than a graphite-lid model, which cools faster than a silicate-lid model.

Our models do not take into account the temperature and pressure effects on the heat capacity of graphite or the pressure effects on thermal conductivity and  viscosity. To assess high-temperature effects, our chosen parameter values are either based on ambient temperature data or measurements in a temperature range relevant to our models depending on the availability of temperature-dependent values of the parameter. Since the interior pressures of planets considered in this study are relatively small, we expect these effects to be small and to not change our conclusions in a qualitative way for planets at least up to the size of Mars. For example, the pressure-dependent silicate viscosity for Kepler-37b-size planets is only 10\% higher than the pressure-independent silicate viscosity. On the other hand, to our knowledge, there are no experimental studies focusing on the effect of high pressure on the properties of graphite. In the future, experimental studies of the high-pressure high-temperature properties of graphite such as thermal conductivity, heat capacity, and activation parameters could further refine assessments of the effect of graphite shells on planetary thermal evolution. 

Since the density of diamond is similar to mantle silicates, diamond is likely to stay mixed with silicates in the mantle, leaving the outer shell to be composed only of low-density graphite. Thus, the maximum possible thickness of the graphite outer shell for a given planet is determined by the graphite-diamond transition pressure above which no graphite exists. As pressures in larger planets increase steeply with depth, the graphite-diamond transition pressure occurs at shallower depths than that for smaller planets. For example, if we assume a temperature-independent graphite-diamond transition pressure of 10~GPa, the maximum possible outer graphite shell thicknesses for a planet with a radius of 3500~km, an Earth-size planet, and a planet twice the size of Earth would be about 1500~km, 400~km, and 100~km, respectively. 

\citet{Unterborn2014} showed that the mixing of diamond with the silicate mantle accelerates the cooling process because of the extremely high thermal conductivity of diamond. If such planets have a graphite outer shell, diamond mixing in the silicate-rich mantle would cool the planet faster while the thermal shielding effect of graphite would slow down the cooling. The net planetary cooling rate of such planets would be faster or slower compared to a lidless planet without graphite or diamond depending on the effect that dominates. For Mercury-size and smaller planets (e.g., Kepler-37b), the mantle pressures would not be high enough to stabilize diamonds.

This study exhibits thermal evolution modeling of carbon-enriched rocky exoplanets that have no solar system analogs. Our calculations show that the overall cooling is greatly affected by the mineralogy of different layers in the planet. As our knowledge of atmospheric and interior composition of rocky exoplanets advances with the data from current and future telescopes (e.g., TESS, CHEOPS, JWST, ELT, PLATO, and ARIEL), the understanding of their interior and surface dynamics also needs to advance with theoretical studies such as this work. 

\begin{acknowledgements}
      We thank Lena Noack for a constructive review that significantly improved this manuscript. We also thank Dan Bower for his valuable comments. This work is part of the Planetary and Exoplanetary Science Network (PEPSci), funded by the Netherlands Organization for Scientific Research (NWO, Project no. 648.001.005). KH also acknowledges financial support from the European Research Council via Consolidator Grant ERC-2017-CoG-771620-EXOKLEIN (awarded to Kevin Heng).
\end{acknowledgements}

%
%

\bibliographystyle{aa}
\bibliography{GraphiteExoplanetsEvolution}

\end{document}